\documentclass[prb,reprint]{revtex4-1} 

\usepackage{amsmath}  
\usepackage{amsfonts} 
\usepackage{graphicx} 
\usepackage{float}

\begin{document}
	
\newcommand{\pderiv}[2]{\frac{\partial}{\partial #1} #2}

\newcommand{\dx}{\mathrm{d}x}
\newcommand{\ds}{\mathrm{d}s}


\title{Topological Effects in Tunneling-Coupled Systems of One-Dimensional Quantum Rings}

\author{Colin J. Riggert}
\email{riggert@ou.edu} 
\affiliation{Homer L. Dodge Department of Physics and Astronomy, The University of Oklahoma, 440 W. Brooks Street, Norman, Oklahoma 73019, USA}
\affiliation{Center for Quantum Research and Technology, The University of Oklahoma, 770 Van Vleet Oval East, Norman, OK 73019}
\author{Kieran J. Mullen}
\affiliation{Homer L. Dodge Department of Physics and Astronomy, The University of Oklahoma, 440 W. Brooks Street, Norman, Oklahoma 73019, USA}
\affiliation{Center for Quantum Research and Technology, The University of Oklahoma, 770 Van Vleet Oval East, Norman, OK 73019}

\date{\today}

\begin{abstract}
Using a model of idealized, crossed one-dimensional quantum wires we construct a novel model for a single electron on tunneling-coupled systems of one-dimensional quantum rings. We explore and find that topology can affect the energetics of the system, and can introduce frustration in the three ring case. We also study the special cases of an external magnetic field that controls the complex phase of the tunneling matrix element, and introduce the knot theory concept of ``writhe" as a new topological quantity for distinguishing one hole systems. We find that writhe not only determines the energetic response of the system to magnetic field strength, but is also responsible for a single particle topological quantum phase transition involving the ground state wave function winding number.

\end{abstract}

\maketitle 

\section{Introduction} 

Since the 1950s, quantum rings have served as a platform for the exploration of quantum behavior, from Aharonov-Bohm (AB) oscillation based interference devices\cite{ABosc1,ABosc2,ABosc3} to the behavior of artificial atoms and molecules.\cite{harrison} In particular, the excluded center and different boundary conditions of rings allows geometries and topologies to be studied that are unobtainable in nature, or even in similar quantum dot setups.\cite{holesMatter}

Previous theoretical work has revealed that, in the idealized case of a one dimensional chain of 1D rings, subject to the Coulomb interaction, this topological exclusion of the center results in a quantum phase transition to an antiferrolectrically polarized state at 0 K.\cite{phaseTransRings} This work, however, considered only the limit of no inter-ring tunneling. Tunnel-coupled systems have also been studied, but in limited cases, such as a single electron on two laterally coupled thick rings, where topology impacts the ``bond strengths" in the resulting artificial molecule.\cite{coupledRings1,coupledRings2,coupledRings3,coupledRings4}

In this paper, we study a model for a single electron  on tunnel-coupled systems of loops using the idealized case of rings as self connected 1D quantum wires. We find that different connections of wire ends to form rings result in topologically distinct systems not accounted for in previous models. We identify the nature of distinctions using knot theory, and numerically examine their effect on system properties both in the absence and presence of a special case of an external magnetic field that controls the complex phase of the tunneling matrix elements. We find that topology affects the energetics of the system as a function of coupling strength, and in the case of an external field, causes a single particle topological phase transition involving the wavefunction winding number.

We note that while some effects, such as the degeneracy of the ground state, are truly topological, the details of others may vary if the rings have different radii. However, for the case where the radii are identical and the rings are tangent we can identify features which are ``paratopological," depending only on connectivity and knottedness.
\section{One-Dimensional Quantum Wires}

The idealized case of an electron on an infinite one-dimensional wire with
an attractive $\delta$-function potential is well known.  We can write the Schr\"odinger equation for such a 
system as
\begin{equation} \label{dim1DDel}
-\frac{\hbar^2}{2m} \frac{\partial^2}{\partial x^2}\psi(x) - \alpha \delta(x) \psi(x) = E \psi(x),
\end{equation}
where $m$ is the mass of the electron and $\alpha$ is the magnitude of the potential. We 
make the substitutions $ x = x_0 s $, $\bar{\alpha} = 2 m x_0\alpha / \hbar^2$, and 
$\epsilon = 2 m x_0^2 E / \hbar^2$, where $s$, $\bar{\alpha}$, and $\epsilon$ are 
dimensionless, and $x_0$ is a characteristic length for the system. With these, we can cast Eq. (\ref{dim1DDel})
in dimensionless variables as
\begin{equation} \label{dimless1DDel}
-\frac{\partial^2}{\partial s^2}\psi(s) - \bar{\alpha} \delta(s) \psi(s) = \epsilon \psi(s).
\end{equation}
The dimensionless energy of the ground state solution\cite{griffithsDelta} to Eq. (\ref{dimless1DDel}) is $\epsilon=-\bar{\alpha}^2/4$, and the normalized wavefunction is $\psi(x) = (\gamma)^{1/2}e^{-\gamma |x|}$, where $\gamma = (-\epsilon/x_0^2)^{1/2}$. The attractive $\delta$-function potential creates a bound state that has an energy lower than any of the system's propagating states, and is exponentially localized about the potential.

\begin{figure}[t!]
	\centering
	\includegraphics[width=\columnwidth]{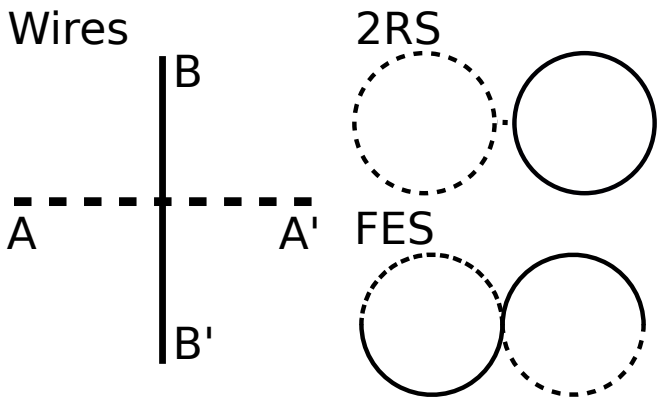}
	\caption{Diagram of our crossed 1D quantum wire model, and the single junction ring systems that can be formed from it by connecting wire ends. The two ring system (2RS) is formed by connecting $A \rightarrow A'$ and $B \rightarrow B'$, and the figure eight system (FES) is formed by connecting $A' \rightarrow B$ and $A \rightarrow B'$. Dashing is used to distinguish the wires when they are connected to form the ring systems.}
	\label{wire_diagrams}
\end{figure}

Next, consider a pair of crossed, finite $(-L\leq x\leq L)$ one-dimensional quantum wires, as shown in Fig. \ref{wire_diagrams}. In this model, we place a single electron on the wire network, and allow it to tunnel between the two wires at their central point of contact. The Hamiltonian for the system is\cite{crossedWires}
\begin{align} \label{dimCrossed}
-\frac{\hbar^2}{2m} \frac{\partial^2}{\partial x^2} \psi(x) + a \delta (x) \phi (0) & = E \psi(x); \\ 
-\frac{\hbar^2}{2m} \frac{\partial^2}{\partial y^2} \phi(y) + a \delta (y) \psi (0) & = E \phi(y),
\end{align}
where $a$ is the strength of the $\delta$-function coupling, $\psi(x)$ is the portion of the electron wavefunction on the horizontal wire, and $\phi(y)$ is the portion of the electron wavefunction on the vertical wire. This system of equations can be solved analytically for the case of zero value boundary conditions on the ends of wires of length $2L$, i.e. $\psi(L) = \psi(-L) = \phi(L) = \phi(-L) = 0$. Making the substitutions  $ x = L s_x $, $y = L s_y$, $\bar{a} = 2 m L a/ \hbar^2$, and 
$\epsilon = 2 m L^2 E / \hbar^2$, we can rewrite the above in dimensionless variables as 
\begin{align} 
\label{dimlessCrossed1}
-\frac{\partial^2}{\partial s_x^2} \psi(s_x) + \bar{a} \delta (s_x) \phi (0) & = \epsilon \psi(s_x); \\ 
 \label{dimlessCrossed2}
-\frac{\partial^2}{\partial s_y^2} \phi(s_y) + \bar{a} \delta (s_y) \psi (0) & = \epsilon \phi(s_y).
\end{align}

Upon inspection, these equations have a similar form to Eq. (\ref{dimless1DDel}). There is again a bound state, with the wavefunction localized around the tunneling contact. Following the analytic solution outlined in Ref.~\onlinecite{crossedWires},  and considering the limit of $\gamma L >> 1$, we arrive at the pair of wavefunctions 
\begin{align}
\label{analyticCrossed1}
\psi(s_x) & = -\frac{\bar{a}}{|\bar{a}|}\phi(0)e^{-\frac{|\bar{a}|}{2} |s_x|}; \\
\label{analyticCrossed2}
\phi(s_y) & = -\frac{\bar{a}}{|\bar{a}|}\psi(0)e^{-\frac{|\bar{a}|}{2} |s_y|},
\end{align}
with dimensionless energy $\epsilon=-\bar{a}^2/4$. The crossed, tunneling-coupled wires have a bound state with exponential localization on each wire around the site of the tunneling, just as in the case of the single wire. Note that while the sign of $\bar{a}$ is arbitrary in Eqs. (\ref{dimlessCrossed1}-\ref{dimlessCrossed2}), it has an effect on the wavefunctions in Eqs. (\ref{analyticCrossed1}-\ref{analyticCrossed2}). A negative value of $\bar{a}$ forces the wavefunctions to be symmetric (i.e. $\psi = \phi$), while a positive value of $\bar{a}$ forces the wavefunctions to be antisymmetric (i.e. $\psi = -\phi$). This can be understood by considering the tunneling terms in Eqs. (\ref{dimlessCrossed1}-\ref{dimlessCrossed2}) as an attractive ``pseudo-potential'' that is felt by the electrons. The sign dependent behavior keeps this pseudo-potential attractive from the perspective of the electron on each wire, allowing for a bound state.

\section{Quantum Rings, $B=0$}
\subsection{Single-Junction Systems}
To obtain a model for a single electron on two tunnel-coupled one-dimensional quantum rings, we have to enforce periodic boundary conditions on the wires, instead of zero value boundary conditions. This is equivalent to connecting the ends of the wires together, creating rings. However, there are two ways this can be done. Using the convention in Fig. \ref{wire_diagrams}, we label the ends of the horizontal wire as $A$ and $A'$, and the ends of the vertical wire as $B$ and $B'$. We can connect $A \rightarrow A'$ and $B \rightarrow B'$, creating a two ring system (2RS), where the tunneling occurs at the rings' point of tangency. We can also connect $A' \rightarrow B$ and $A \rightarrow B'$ (or, equivalently, $A' \rightarrow B'$ and $A \rightarrow B$). This creates a system that is essentially one large ring twisted on itself to make a figure-eight system (FES), with the tunneling occurring where the system crosses itself (Fig. \ref{wire_diagrams}). While both appear to have two rings, only one of them truly does, so we shall refer to them as single junction systems instead.

\begin{figure}[b!]
	\includegraphics[width=\columnwidth]{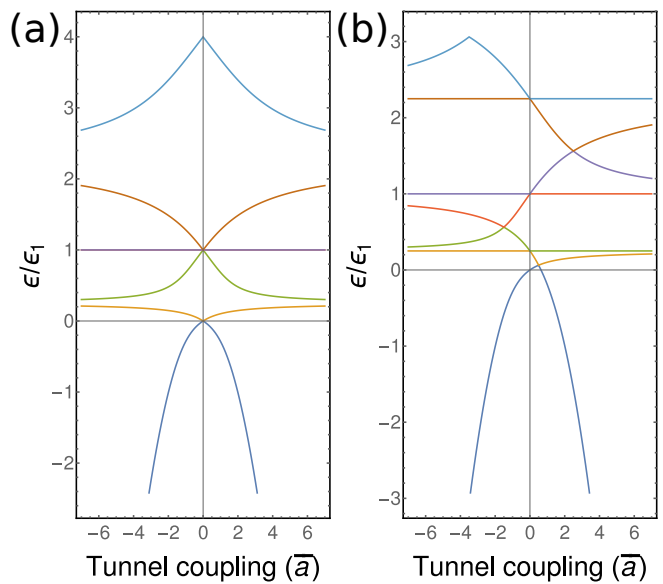}
	\caption{Energy spectra for a single electron on various single junction system topologies as a function of dimensionless tunneling strength $\bar{a}$. (a) is the spectrum for the two ring system, and (b) is the spectrum for the figure eight system. Energy is scaled in units of $\epsilon_1=\hbar^2/2mR^2$, the energy of the first excited state of a ring with radius $R$.}
	\label{single_junction_specs}
\end{figure}

\begin{figure*}[t!]
	\centering
	\includegraphics[width=\textwidth]{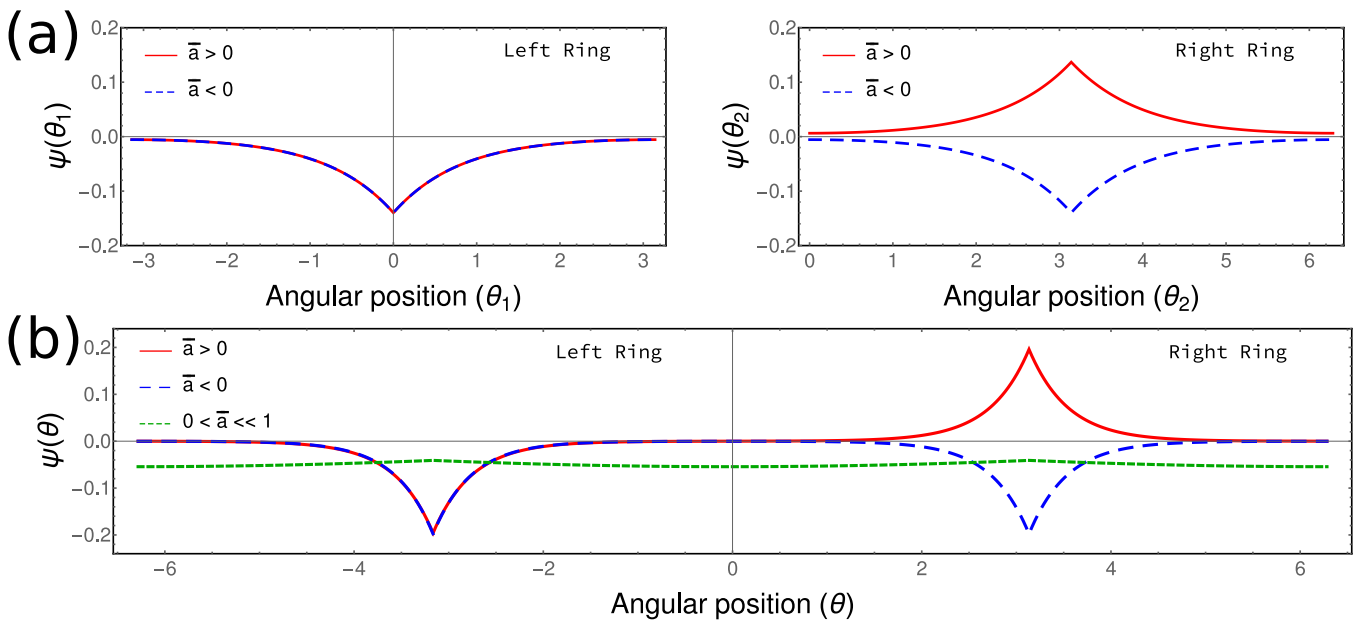}
	\caption{Plots of wavefunctions for a single electron on (a) the two ring system and (b) the figure eight system for different coupling regimes. In both cases, strong negative (positive) coupling introduces (anti)symmetric localization. Small positive coupling in the figure eight results in a non-localized state due to competing gradient energy costs associated with antisymmetry.}
	\label{single_junction_wavefunctions}
\end{figure*}

These systems are topologically distinct: the 2RS has two holes, and the FES has one. As a result, the energy spectra of these systems will be different, even in the absence of tunneling. In the idealized 1D limit, the energy levels of a particle on a single ring of constant magnitude curvature go as $m^2 E_0$, where $m$ is an integer, and $E_0$ is an energy dependent on the size of the ring.\cite{griffithsRings,shankarRings} Thus, the ground state of the ring, where $m=0$, is a true zero-energy state, regardless of the size of the ring. Since the 2RS and FES have constant curvature, the ground states for both the 2RS and FES have zero energy in the absence of tunneling. However, because there is no tunneling, an electron can only exist on one ring at a time, and thus the two systems have different degeneracies for this ground state. The 2RS has two rings, and is thus doubly degenerate, while the one ring FES has a unique ground state. Generalizing these results, these sorts of ring systems have $n$-fold degenerate ground states in the absence of tunneling, where $n$ is the number of holes the system has, but these ground states all have zero energy, regardless of $n$.

To find the spectra of these systems for arbitrary tunneling strength, we adapt the Hamiltonian in Eqs. (\ref{dimlessCrossed1}-\ref{dimlessCrossed2}) to enforce periodic boundary conditions for the 2RS or FES. We then numerically solve the problem by discretizing angular displacement on each ring on a one-dimensional mesh. Using the framework of this numerical solution, we vary the value of the tunnel coupling element $\bar{a}$ and create a plot of energy as a function of $\bar{a}$. This resulting energy spectrum is plotted for both the FES and 2RS in Fig. \ref{single_junction_specs}.

In both spectra, when $\bar{a}=0$, we see the single or double degeneracy in the ground state arising from the topological differences in hole counts between the two systems. We also see that, while the 2RS spectra is symmetric about the $\bar{a}=0$ axis, the FES spectra is not, with an asymmetry for small positive values of $\bar{a}$. This is due to the symmetry behavior of the ground state wavefunction. In the crossed, disjoint wires case, when $\bar{a}<0$, the wave functions on the two wires are symmetric, but when $\bar{a}>0$, the wavefunction on the two wires is antisymmetric. When the wires are connected on themselves to form the 2RS, the wires are still disjoint sets, and so this behavior is unchanged, as shown in Fig. \ref{single_junction_wavefunctions}.

 In contrast, the FES wavefunction feels a phase twist, increasing its kinetic energy. For small positive values of $\bar{a}$, this gradient energy cost is larger than the energy saved by adopting the antisymmetric ground state configuration, and the wavefunction instead adopts a symmetric configuration that is weakly repelled from the tunneling site. As a result, unlike the crossed wires of Eqs. (\ref{analyticCrossed1}-\ref{analyticCrossed2}), the energy of this system depends upon the sign $\bar{a}$. For negative and large positive values of $\bar{a}$ the wavefunction behavior mirrors that of the 2RS system. The full wavefunction behavior of the FES is shown in Fig. \ref{single_junction_wavefunctions}. This gradient energy asymmetry is an effect solely of the topology of the FES system, and explains the asymmetry of the calculated energy spectra for the FES at small positive values of $\bar{a}$. 
 
 In the zero-coupling limit, excited states on rings are admitted which have nodes at both tunneling sites. As such, the tunneling term in Eqs. (\ref{dimlessCrossed1}-\ref{dimlessCrossed2}) disappears, and these states, and their energies, are unaffected by tunneling. These states explain the flat lines in the spectra seen in Fig. \ref{single_junction_specs}. Because the FES is effectively a larger ring, the energies of these states on the FES are lower, resulting in the increased density of such states in the FES spectrum as compared to the 2RS spectrum.
 
 We have treated the sign of the tunneling coupling as a controllable parameter, but it may be argued that generically it is negative, since an electron can lower its energy by delocalizing across the intersection. However, this can produce topological effects in the higher energy states which have nodes and therefore change sign as they wind about the system.  In general, no matter the sign of $\bar{a}$, we can find states where topology of the structure alters the energy.

\begin{figure}[t!]
	\includegraphics[width=\columnwidth]{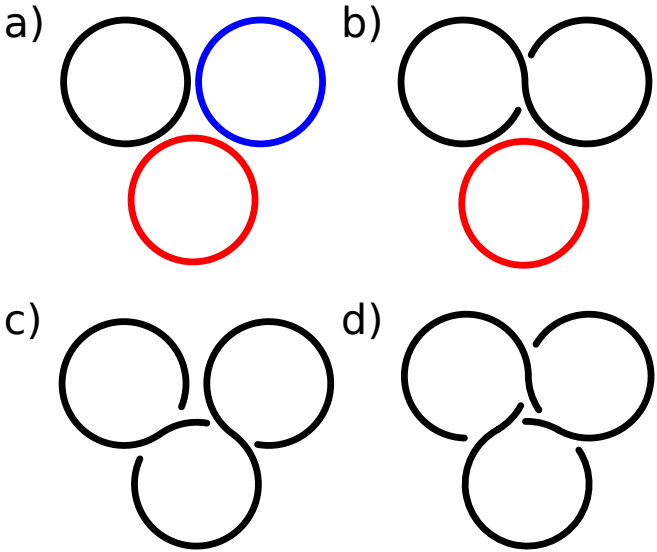}
	\caption{Possible system topologies for triple junction systems. Dashing is used to distinguish disjoint wire elements. Wire crossing tunneling is indicated by points where rings touch, and gaps indicate point of nearest approach tunneling. The topologies illustrated are the (a) three ring system, (b) the two hole system, (c) the mouse ear system, (d) and the triple crossing system.}
	\label{triple_junction_drawings}
\end{figure}

\subsection{Triple-Junction Systems}
Triple ring arrays are potentially richer, because they can exhibit frustration.\cite{frustration} For positive values of $\bar{a}$, the wavefunction on each ring tries to adopt a phase shift of $\pi$ with respect to its nearest neighbor, but the wavefunctions on our three rings cannot all simultaneously satisfy this antisymmetry. The ground state for negative $\bar{a}$ does not have such an antisymmetry requirement, but in the first and second excited states, we admit a node into the wavefunction that again forces a phase shift between adjacent rings, re-introducing frustration. In all systems below, this frustration creates  degeneracy or near-degeneracy between the ground and first excited (first and second excited) states in states with sufficiently positive (negative) $\bar{a}$.

\begin{figure}[t]
	\includegraphics[width=\columnwidth]{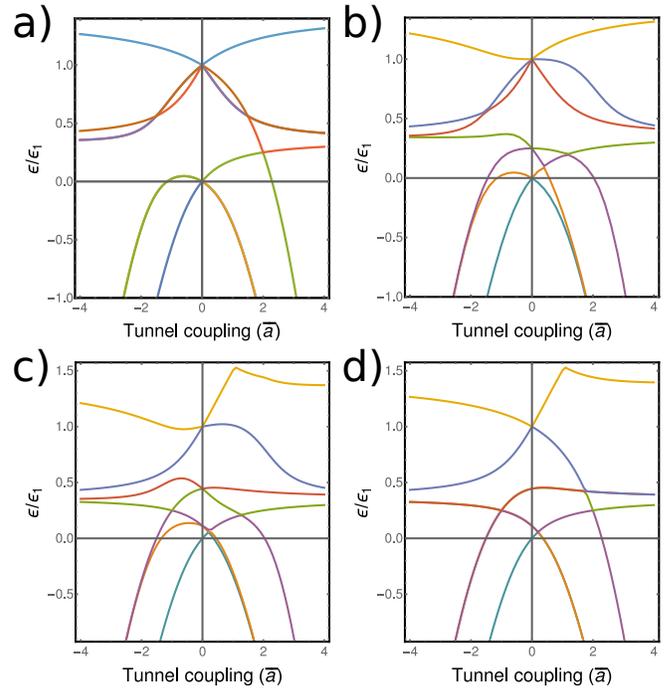}
	\caption{Energy spectra for a single electron on various triple junction system topologies as a function of dimensionless tunneling strength $\bar{a}$. Figs. (a)-(d) correspond to topologies (a)-(d) of fig.\ref{triple_junction_drawings}. Energy is scaled in units of $\epsilon_1=\hbar^2/2mR^2$, the energy of the first excited state of a ring with radius $R$.}
	\label{triple_junction_specs}
\end{figure}

There are several topologically distinct starting $\bar{a}=0$ geometries. The first is one that has three disjoint rings (3RS), as shown in Fig. \ref{triple_junction_drawings}a. The 3RS Hamiltonian is similar to Eqs. (\ref{dimlessCrossed1}-\ref{dimlessCrossed2}), with additional tunneling terms added to reflect system geometry. The numerically calculated spectra as a function of $\bar{a}$ is plotted in Fig. \ref{triple_junction_specs}a. Because the 3RS has three holes, the ground state at $\bar{a}=0$ is triply degenerate. However, because of frustration, we do not observe symmetry in the spectra, even though the rings form disjoint sets.
We can devise other topologically distinct triple-junction systems. The two-hole system (2HS), made from the combination of a FES and a disjoint ring, is shown in Fig. \ref{triple_junction_drawings}b. The Hamiltonian for this system is similar to that of the 3RS, but with the periodic boundary conditions altered for the FES. The eigenenergies are plotted in Fig. \ref{triple_junction_specs}b. The ground state at $\bar{a}=0$ is now doubly-degenerate, reflecting the two hole system. The spectra are not symmetric about $\bar{a}$ due to the gradient energies in the FES, and we see the expected frustration degeneracies.

Further reducing the hole count, there are two ways to create a triple-junction system with one hole. In the first, all tunneling junctions correspond to the wire crossing itself, creating a triple-crossing system (TCS) (Fig. \ref{triple_junction_drawings}d). In the second, two of the junctions correspond to the wire crossing itself, but the third corresponds only to wire tangency, creating a mouse ear system (MES) (Fig. \ref{triple_junction_drawings}c).

The Hamiltonians of these systems are similar to that of the 3RS or 2HS, with differently enforced periodic boundary conditions due to only containing one disjoint set. However, because the wire is ``wrapped" differently between the TCS and MES, when we discretize angular position on a 1D mesh, distance between tunneling location differs in the two systems. This leads to differences in the spectra as a function of $\bar{a}$, shown in Fig. \ref{triple_junction_specs}c-d. Both spectra display a unique ground state at $\bar{a}=0$ (due to only having one hole), the asymmetry arising from ground state gradient energies, and the degeneracies from frustration. However, despite both having the same hole-count, the spectra for the TCS and MES are not identical. This suggests the need for an additional topological quantity to distinguish the TCS and MES, and to better characterize these reduced hole systems. To explore this second topological quantity, we consider our triple-junction systems in the presence of external magnetic fields. 
\section{Quantum Rings, $B \neq 0$}
\subsection{Knot Theory and the AB Effect}

\begin{figure}[t]
	\centering
	\includegraphics[width=\columnwidth]{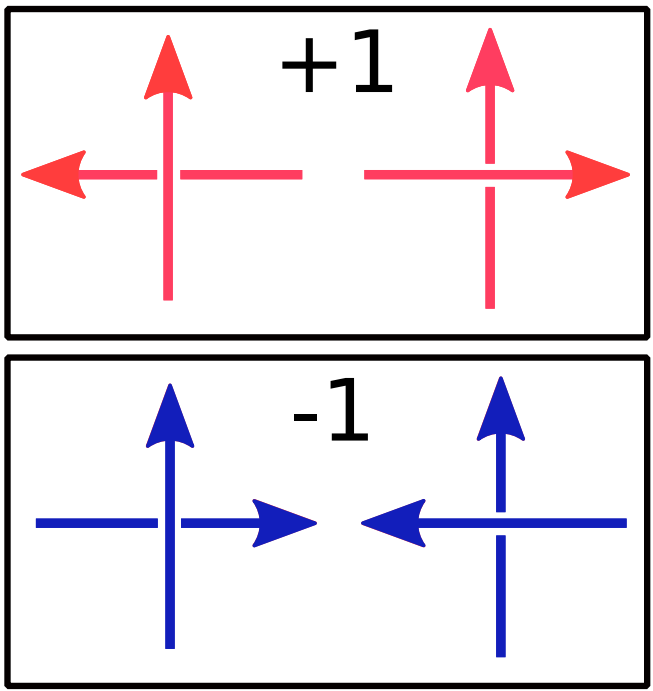}
	\caption{Possible oriented crossings and their contributions to writhe. Each crossing in a knot is assigned a value of $\pm1$ based on which of the above crossings it matches, and the total writhe is found by summing these values over all crossings in the system.}
	\label{writhe}
\end{figure}

Specifically, in order to further explore the role of topology in these systems we investigate the effect of adding a complex phase to the tunneling constant $\bar{a}$, such that $\bar{a} \rightarrow \bar{a}e^{i\phi}$. Physically this might be accomplished by  applying an in-plane magnetic field, so that it has no effect on the in-plane motion of the electrons, but introduces an Aharonov-Bohm (AB)\cite{ABeffect,ABevidence1,ABevidence2} phase, $e^{i\phi}$, when an electron moves out-of-plane, tunneling from one wire to the next. The phase $\phi$ is given by
\begin{equation} \label{ABEffect}
\phi = -\frac{e}{\hbar c} \int_P \vec{A}\cdotp d\vec{l},
\end{equation}
where $P$ is the path, $\vec{A}$ is the vector potential, and $e$ is the magnitude of the electron charge.

\begin{figure}[t!]
	\centering
	\includegraphics[width=\columnwidth]{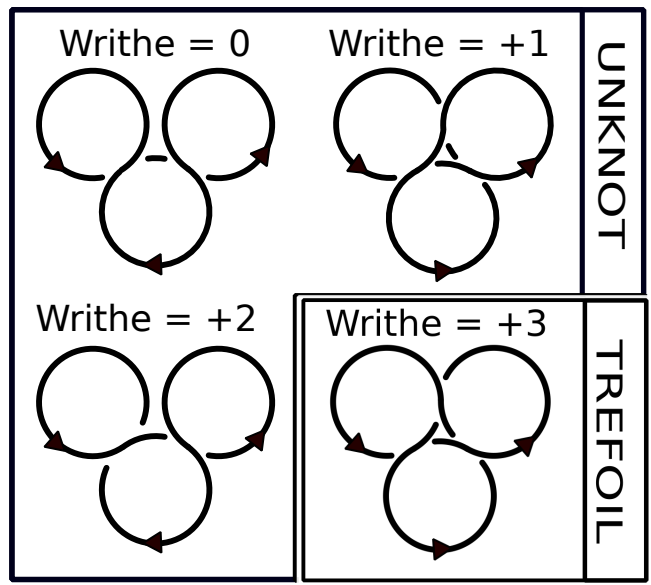}
	\caption{Diagram showing the four distinct one hole triple knotted systems, grouped by knottedness and labeled by writhe. Three such systems are equivalent to the canonical unknot. One is truly knotted, and equivalent to the trefoil.}
	\label{one_hole_knots}
\end{figure}

However, we need to be careful about the ``over'' and ``under''-ness of the crossings, as whether the electron jumps from the top wire to the bottom or vice versa at a crossing flips the sign of the induced phase $\phi$. Thus, the system is sensitive to the \textit{total} ``over/under-ness" of the crossings. This global system property is reflected in the topological quantity of ``writhe" from knot theory.\cite{writhe} Writhe is calculated by orienting a knot and assigning each oriented crossing the associated values from Fig. \ref{writhe}. The writhe is the sum of these values over all system crossings.

We can further apply knot theory with the concept of ``knottedness," which describes the simplest knot to which a system can be unwound.\cite{knotedness} Some one hole systems can be unwound to the ``unknot" - a simple circle - while others are irreducibly knotted.
\subsection{Triple Junction Knots}
Writhe and knottedness allow us to identify four topologically distinct one hole, triple ring knotted systems, illustrated in Fig. \ref{one_hole_knots}. Three of these systems are equivalent to the ''unknot," or simple loop, and one is truly knotted - equivalent to the trefoil. All four have different writhes, taking the integer values between 0 and 3, inclusive.

\begin{figure}[t!]
	\centering
	\includegraphics[width=\columnwidth]{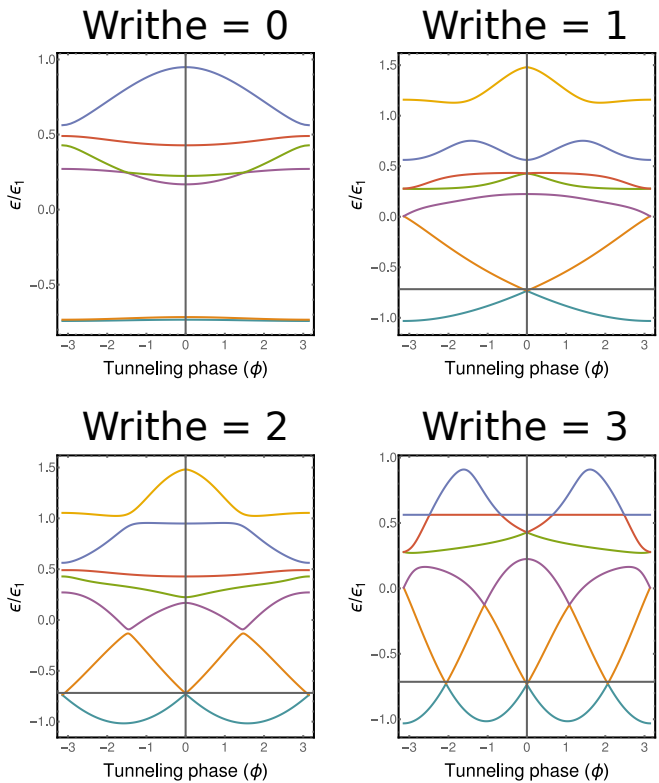}
	\caption{Energy spectra for a single electron in the four possible one hole knotted triple ring systems, labeled by writhe. For each system, energy is plotted as a function of the tunneling phase shift $\phi$ with a tunnel coupling strength of $\bar{a}=1.5$. Energy is scaled in units of $\epsilon_1=\hbar^2/2mR^2$, the energy of the first excited state of a ring with radius $R$.}
	\label{one_hole_specs}
\end{figure}

The writhe of a knotted one hole, triple ring system determines the energetics of the system when placed in the specially oriented magnetic field described above. To see this, for each system, we keep the tunnel coupling magnitude constant, and vary the tunneling phase, $\phi$, from $-\pi$ to $\pi$. For a coupling magnitude of $\bar{a}=1.5$, the resulting spectra are shown in Fig. \ref{one_hole_specs}. In each spectra, the number of level crossings between the ground state and first excited state in one $2\pi$ period of tunneling phase matches the writhe.

This can be understood as an effect of magnetic phase commensuration. Tunneling creates multiple possible self connected paths through the system for any starting point, and this creates interference effects, since each path picks up different amount of tunneling phase. Level crossings occur when these path-dependent accumulated phases interfere constructively.

When crossings have opposite writhe, tunneling through them results in phases which cancel each other out. When crossings are oriented the same way, the phase picked up in traversing them accumulates as. Thus, systems with more identically oriented crossings can obtain constructive interference (and thus a level crossing) with a smaller applied field. Since the magnitude of the writhe describes the degree to which the crossings in a system are all oriented in the same way, systems with higher writhe display more frequent level crossings, as observed in Fig. \ref{one_hole_specs}.

\begin{figure}[t]
	\centering
	\includegraphics[width=\columnwidth]{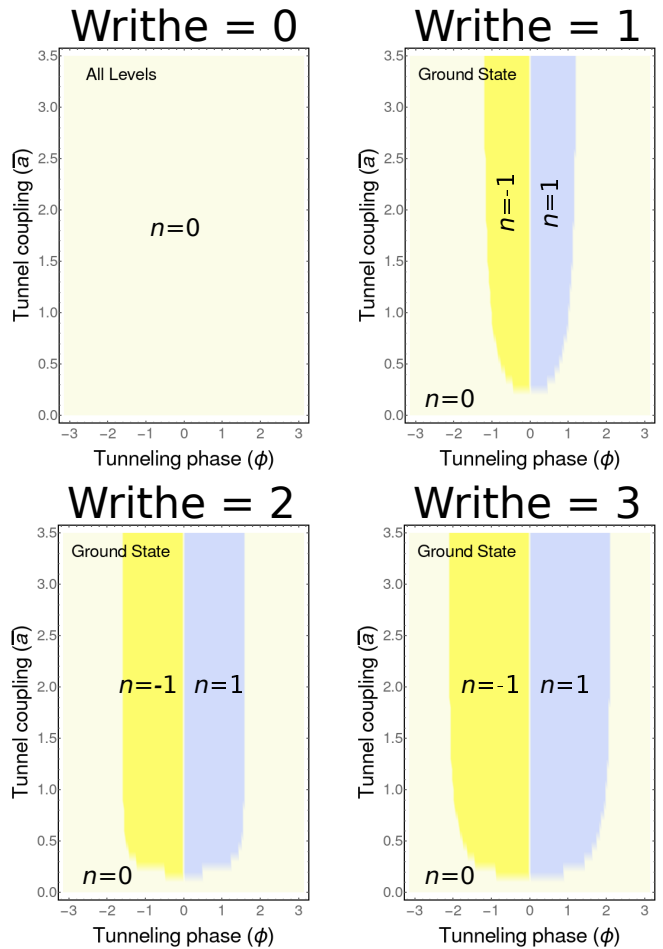}
	\caption{Phase diagram showing the winding number, $n$, of the ground state wavefunction in knotted one hole systems as a function of $\bar{a}$ and $\phi$. All systems with nonzero writhe experience a phase transition with spontaneous changes in winding number as boundaries in parameter space are crossed. The location of these boundaries in parameter space depends on the system writhe. The system with zero writhe experiences no phase transition, indicating that this is a topological phase transition.}
	\label{phaseTransition}
\end{figure}

Placing our system in this specially oriented magnetic field also allows us to observe a topological quantum phase transition. For any system with nonzero writhe, by keeping the tunneling strength constant and sweeping the tunneling phase, we can cause the winding number, $n$, of the ground state wavefunction to spontaneously change. This winding number is calculated as 
\begin{equation}
n=-\frac{i}{2\pi}\int z^*\pderiv{s}{(z)}\ds ,
\end{equation}
where $s$ is a parameter for our position on the system, and
$z\equiv\psi/|\psi|$
is a mapping of our wavefunction on to $S^1$. As the tunneling phase is swept, it becomes energetically favorable for the wavefunction to accommodate the changing phase contributions by changing how it ``twists" in the complex plane as it goes about the system, resulting in a change in winding number.

As the winding number is a global property that is changing spontaneously as system parameters are altered, this phenomena is a quantum phase transition.\cite{sachdev} The phase diagrams for the various one whole knots are shown in Fig. \ref{phaseTransition}. Further, because the transition only occurs in systems with nonzero writhe, and because the system writhe effects the location and curvature of the boundaries between regions in phase space, this transition is a result of system topology. It can thus be understood as a single particle topological phase transition.

\section{Conclusion}
We have examined the properties of tunnel coupled one-dimensional quantum rings in the two and three rings case, using a model of tunnel coupled quantum wires. We showed system topology plays a large role in determining these properties. In all cases, the hole count of the system effects degeneracy, symmetry, and general form of the energy sp\textsl{}ectra found when the tunnel coupling strength, $\bar{a}$, is varied. Furthermore, in the case of triple ring systems, we have shown that the system energetics display behavior that indicates frustration in a asingle particle system. This frustration is a result of an antisymmetry between nearest neighbors in the ground state wavefunction that occurs for positively coupled systems. While hole count did distinguish system with different spectra in the three ring case, it did not do so completely, as we were able to construct two systems that had different spectra but both had one hole.

To further distinguish these systems, we borrowed the concepts of ``knottedness" and ``writhe" from knot theory, and found that these additional global descriptors allowed us to fully topologically distinguish our systems. We have also shown that when these knotted one hole systems are placed in a magnetic field that causes the electron to pick up a phase in tunneling, the writhe determines the number of level crossings between the ground and first excited states as the tunneling phase is varied by one full period. Finally, we demonstrated that the system topology described by the writhe is also responsible for a topological quantum phase transition involving the winding number of the ground state wave function.

That system topology plays such a large role in determining properties for small ring count systems suggests several avenues for future work. One could extend this sort of search to large rings of lattices with periodic boundary conditions, or create lattices with knots at each lattice site. One could also further explore the frustration displaced in the triple ring systems, and build ring systems in more geometrically frustrated configurations, such as the Kagome lattice.

These results also suggest a richness in the many body case. It is known that tunneling-free ring arrays display a quantum phase transition due to the Coulomb interaction,\cite{phaseTransRings} and the inclusion of tunneling could further add depth to this transition, or result in entirely new system phases or system behavior, as system topology could lead to exotic sorts of topological phase transitions or collective excitations.

\end{document}